\documentclass[]{spie}  

\renewcommand{\baselinestretch}{1.0} 
 
\usepackage{amsmath,amsfonts,amssymb}
\usepackage{graphicx}
\usepackage[colorlinks=true, allcolors=blue]{hyperref}
\usepackage[flushleft]{threeparttable} 
\usepackage{booktabs}
\usepackage{array}
\usepackage{footmisc}

\begin{document} 

\title{Wide-field ultra-narrow-bandpass imaging with the Dragonfly Telephoto Array}

\author[a,b]{Deborah M. Lokhorst}
\author[a,b]{Roberto G. Abraham}
\author[c]{Pieter van Dokkum}
\author[a,b]{Seery Chen}
\affil[a]{David A. Dunlap Department of Astronomy \& Astrophysics, University of Toronto, Canada}
\affil[b]{Dunlap Institute for Astrophysics, University of Toronto, Canada}
\affil[c]{Department of of Astronomy, Yale University, USA}

\authorinfo{Further author information: (Send correspondence to D.M.L.)\\D.M.L.: E-mail: lokhorst@astro.utoronto.ca}

\pagestyle{empty} 
\setcounter{page}{301} 
 
\maketitle

\begin{abstract}

We describe plans for adding a wide-field narrow-band imaging capability to the Dragonfly Telephoto Array. Our plans focus on the development of the `Dragonfly Filter-Tilter', a device which places ultra-narrow bandpass interference filters ($\Delta\lambda \approx 1$ nm) \textit{in front} of each of the lenses that make up the array. The filters are at the entrance pupil of the optical system, rather than in a converging beam, so their performance is not degraded by a converging light cone. This allows Dragonfly to image with a spectral bandpass that is an order of magnitude narrower than that of telescopes using conventional narrow-band filters, resulting in a large increase in the contrast and detectability of extended low surface brightness line emission.  By tilting the filters, the central wavelength of the transmission curve can be tuned over a range of 7 nm, corresponding to a physical distance range of about 20 Mpc of extragalactic targets. A further benefit of our approach is that it allows off-band observations to be obtained at the same time as on-band observations, so systematic errors introduced by rapid sky variability can be removed with high precision. Taken together, these characteristics should give our imaging system the ability to detect extremely faint low-surface brightness line emission. Future versions of the Dragonfly Telephoto Array may have the sensitivity needed to directly image the circumgalactic medium of local galaxies.  In this paper, we provide a detailed description of the concept, and present laboratory measurements that are used to verify the key ideas behind the instrument. 
\end{abstract}

\keywords{narrow-band imaging; wide-field imaging; low surface brightness galaxies; instrumentation}

\section{INTRODUCTION}
\label{sec:intro}  


Astronomers have been undertaking wide-area emission line surveys of the sky for many decades (see, e.g., Ref.~\citenum{mart19} for a review), and they have adopted a wide range of techniques.
There is, of course, no `best' technique, since the right approach depends on the properties of the targets and on the science goals of the observer. 
For example, slitless spectroscopy can be used to obtain spectra for many objects simultaneously, but this technique is only effective for studying bright sources where sky contamination is not a limiting factor. For faint targets, multi-slit or multi-fiber spectroscopy is generally a better choice, although both approaches are best used to investigate compact targets, whose projected sizes are comparable to the slit width or to the projected extent of the fiber (or fiber bundle) at the focal plane. At the other extreme, techniques used for probing individual objects are generally not well suited to probing line emission from very extended structures. This situation may now be changing, because of the impressive capabilities of high-performance integral field spectrometers, such as the Keck Cosmic Web Imager on the Keck Observatory\cite{morr18} and the Multi-Unit Spectroscopic Explorer on the European Southern Observatory's Very Large Telescope (VLT)\cite{baco10}. However, at present these are limited to small fields-of-view ($<$ 1 arcmin$^2$), and their complicated optical configurations and many optical surfaces can limit their low surface brightness performance. 


As a result of these considerations, narrow band imaging is the method of choice for obtaining low-resolution spectral maps down to faint surface brightness limits over very wide areas of the sky. However, narrow-band imaging has a number of disadvantages:

\begin{enumerate}
\item The spectral resolving power, $\mathcal{R} = \lambda/\Delta\lambda$, of narrow-band filters is generally poor. On most large telescopes, narrow-band interference filters\footnote{At present there are no good alternatives to interference filters for narrow-band imaging with large telescopes. Colored glass filters can only be manufactured with relatively broad bandpasses, and the transmissivity of colored glass depends upon its thickness. Filters that are rigid enough to be large must also be quite thick, which limits their performance.} have a bandwidth that is $\sim$10 nm wide, corresponding to a spectral resolving power of about $R \approx 65$ at the wavelength of ${\rm H}\alpha$ (see Table~1 for a compendium of the properties of narrow-band filters available on large telescopes). There are many applications where narrower bandwidths are desirable, either for science reasons (such as distinguishing between H$\alpha$ at 656.3 nm and [NII] at 658.3 nm), or for operational reasons (such as avoiding contamination by night sky lines that can overwhelm the signal from the object of interest at low surface brightness levels). It is possible to manufacture filters with bandpasses $\ll10$ nm in quite large sizes\footnote{For example, Hyper SuprimeCam on the Subaru telescope uses interference filters that are 580mm in diameter. Large narrow-band interference filters are certainly challenging to manufacture, and both the difficulty and expense increase as the bandpass decreases. They do exist though.}, but they have not been considered useful, for reasons described next.
\item The performance of narrow-band interference filters degrades in fast light cones, and varies with field angle. This is because the filters rely on constructive and destructive interference of the wave-front, so their efficiency depends on the optical path length through the coating stack.  As beams get faster, the central wavelength shifts, peak efficiency drops, and performance becomes a strong function of position across the field (see the discussion in Section~\ref{sec:InstrumentDesign}). These limitations increase in severity as the bandpass of the filter is narrowed. This poor performance in fast beams, rather than manufacturability, is the factor limiting the effective bandwidth of narrow-band filters on large telescopes\footnote{All telescopes must conserve \'etendue, so a large telescope with a monolithic aperture must also have long focal length, and the focal length sets the image scale at the focus. Therefore, unless their fields of view are tiny, or the sizes of their focal planes are enormous, large telescopes must use very fast optics in their re-imaging cameras.}.
\item A separate filter is required for each wavelength channel of interest. Since large interference filters are costly to manufacture, it can be difficult to use narrow-band filters to probe a significant range of wavelength space without painful trade-offs between the desired spectral resolution and the total cost of the filter set. This is particularly problematic for extragalactic sources, as the emission lines are redshifted.
\end{enumerate}

In the present paper we show that the limitations described above are not fundamental, and describe how filters with extraordinarily narrow bandpasses ($\Delta\lambda\lesssim1$ nm) can be used in large telescopes operating in
optically fast imaging configurations, {\em with no degradation in performance}. This is achieved by: (1) using the mosaic telescope concept to split a telescope's aperture into multiple small optical elements whose images are digitally combined to yield the equivalent of an image obtained through a much larger seeing-limited telescope; (2) placing very narrow bandpass interference filters {\em in front} of the optical elements that define the entrance pupil of the telescope, so they are in a perfectly collimated beam with a focal-ratio of infinity; and (3) devising a mechanism to tilt individual filters, thereby allowing them to be tuned over a range of wavelengths. 

To test our ideas, we have built a small narrow-band imaging array as an extension of the Dragonfly Telephoto Array (hereafter, Dragonfly; Ref.~\citenum{abra14}). Individual Dragonfly lenses are small enough (143 mm aperture in the current telescope) that interference filters can be readily manufactured in the required sizes, and mechanisms to tilt the filters are straightforward to implement. At the same time, individual lenses are large enough to make the telescope seeing-limited at the observing site.  By growing the array gradually, we aim to eventually achieve the effective aperture and sensitivity needed to reach a surface brightness limit of $\sim$1000 ph s$^{-1}$ cm$^{-2}$ sr$^{-1}$, which is the limit needed to directly image H$\alpha$ emission from the circumgalactic medium of local galaxies\cite{lokh19}. The bandpass of the system is adjustable, so the upgraded Dragonfly array will not only be a wide-area ultra-narrow-band imager, it will also be a wide-angle tunable imaging spectrometer capable of mapping line emission out to distances well beyond the Virgo cluster.

\begin{table*}[ht]
\caption{Compendium of current narrowband filters available on large telescopes including the Canada France Hawaii Telescope (CFHT) MegaPrime wide-field optical imaging facility, the Gemini Multi-Object Spectrograph (GMOS) on the Gemini Observatory, the Hyper Suprime-Cam (HSC) on the Subaru Telescope, the Spectral Camera at the Las Cumbres Observatory, the   SOAR Optical Imager (SOI) on the Southern Astrophysical Research Telescope, as well as the Kitt Peak National Observatory (KPNO) imagers, such as the Burrell Schmidt.} 
\label{tab:FilterCompendium}
\begin{center}       
\begin{threeparttable}
\begin{tabular}{ *5l }    
\toprule
Telescope &	Central Wavelength & Bandpass & Peak Transmission\tnote{a} & Name \\
 &	 (nm) &  (nm) &  (\%) &  \\\midrule
CFHT MegaPrime &	659         &   10.4        &   93	            & Ha \\
            &   	671.8       &   10.7	    &   95              & HaOFF \\
            &   	500.6	    &   9.9         &   91	            & OIII \\
            &   	510.5	    &   9.5	        &   93	            & OIIIOFF \\
	        &       395.1	    &   9.6	        &   91	            & CaHK \\\midrule
Gemini GMOS	&       468	        &   8		    &     -             & HeII \\
	        &       478	        &   8		    &     -             & HeIIC \\
	        &       499         &   5	        &     -             & OIII \\
            &   	514	        &   10		    &  -                & OIIIC \\
	        &       656	        &   7           &  -             	& Ha \\
	        &       662	        &   6		    &  -                & HaC \\
	        &       672	        &   4.3		    &  -                & SII \\
	        &       679	        &   4.8		    &  -                & OVIC \\
	        &       684	        &   4.9		    &  -                & OVI  \\\midrule
Subaru HSC	&       502.9	    &   7.5		    &  -                & [OIII] \\
        	&       659.6	    &   9.9		    &  -                & Halpha \\
	        &       704.6	    &   10		    &  -                & Barr 7040A \\
	        &       712.6	    &   7.3		    &  -                & Asahi 7110A \\
	        &       815.4	    &   11.9		&   -               & Barr 8160A \\
            &   	913.9	    &   12.6		&   -               & Asah 9210A \\
	        &       919.6       &	13.2		&   -               & Barr 9210A \\\midrule
Las Cumbres Spectral	& 500.7	&   3		    &   -               & OIII \\
	        &       515	        &   15		    &   -               & D51 \\
	        &       656.3	    &   30		    &-                  &  H-alpha \\ 
	        &       486	        &   30		    &-                  &  H-beta \\ 
	        &       383	        &   6		    & -                 & Skymapper v \\\midrule
SOAR Optical Imager	&   656.3	&   7.5		    &  -                & Halpha  \\
            &   	673.8	    &   5		    &   -               & [S II]  \\
	        &       660	        &   7.5		    &    -              & Halpha 6600  \\
	        &       501.9       &	5	        &     -             &	[OIII]  \\
	        &       777.9       &	12.5		&      -            & TiO Wing \\ 
	        &       812.3	    &   11.6		&       -           & CN Wing \\\midrule
KPNO\tnote{b} &     500.8       &   4.5	        &   84	            & [O III] \\
            &	    502.2	    &   5.3	        &   83	            & [O III] redshifted \\
	        &       487.1	    &   6.3	        &   79	            & HBeta \\
	        &       487.6	    &   2.5 	    &   79	            & H-beta narrow \\
	        &       656	        &   2.9	        &   80	            &   H-alpha narrow \\
            &   	657.3	    &   6.7         &	83              &	H-alpha\_2 \\
	        &       661.8       &	7.4	        &   79	            &   H-alpha6\_2 \\ 
	        &       665.3	    &   6.8	        &   83	            &   H-alpha9\_2 \\ 
	        &       670.9	    &   7.1	        &   83	            &   H-alpha14\_2 \\
	        &       674.6	    &   8.3	        &   67	            &   H-alpha +18nm \\ 
            &   	678.6	    &   7.8	        &   65	            & H-alpha +22nm \\
	        &       683.2	    &   7.5	        &   67	            & H-alpha +27nm \\ 
	        &       687.7       &	7.6	        &   85	            & H-alpha +31nm \\
\bottomrule
\end{tabular}
  \begin{tablenotes}
  \item[a] Missing for filters where not explicitly stated on observatory website.
  \item[b] A selection of the 4-inch filters available for use on telescopes at the KPNO facility.
  \end{tablenotes}
\end{threeparttable}
\end{center}
\end{table*}

\section{CONCEPT}
\label{sec:InstrumentDesign}
   
The central wavelength of the transmission function of interference filters depends on the the angle of incidence of the incoming light cone. Most interference filters can be modelled as multi-cavity transmissive materials, where the index of refraction for each cavity is independently included in the model.  For a single cavity filter, and with the assumption of small angles, the shifted wavelength, $\lambda '$, from the incident wavelength, $\lambda$, is given as a function of incident angle $\theta$ by the following equation:
\begin{equation}
\label{eq:shift}
\lambda ' = \lambda \sqrt{1 - (\frac{1}{n_{c}}~\mathrm{sin}~\theta)^2}
\end{equation}
where $n_{c}$ is the index of refraction of the cavity medium\cite{smit00}. The expression describing the dependence of wavelength on incident angle for multi-cavity filters (or multi-layer films) is more complex, but it simplifies to Equation~(1) for small angles. As such, we can approximate the behaviour of the narrow-band filters using the above equation where the index of refraction is replaced by an `effective' index of refraction for the stack of layers in the filter.

\begin{figure} [hbt]
   \begin{center}
   \begin{tabular}{c} 
   \includegraphics[height=7.1cm]{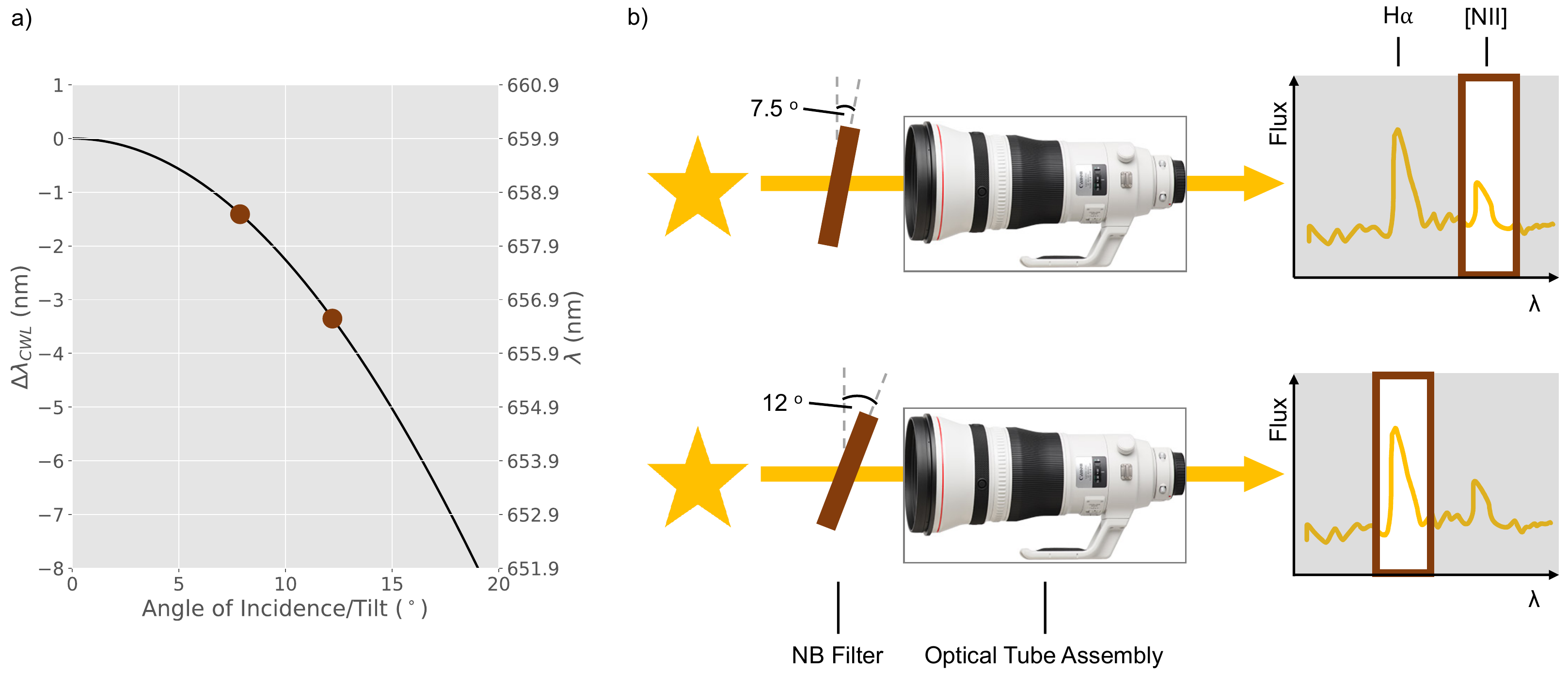}
   \end{tabular}
   \end{center}
   \caption[tiltshift1] { \label{fig:tiltshift1}  (a) The predicted shift of the central wavelength of the filter as a function of angle of incidence (or, equivalently, the tilt of the filter with respect to normal incidence) for the dielectric filter recipe described in the text. (b) A schematic illustration showing how tilting the filters at the entrance pupil shifts the central wavelength of the filter transmission function from the [NII] line (at a $7.5^\circ$ tilt) to the H$\alpha$ line (at a $12^\circ$ tilt).
   }
\end{figure}  

Understanding the implications of Equation~(1) for narrow-band astronomical filters is best done by considering a specific example. Detailed recipes for the evaporated dielectric stacks of commercially available interference filters are not generally disclosed by manufacturers, but we can take an educated guess for the effective index refraction by using the indices of refraction of materials commonly used in optical filters as a starting point. For example, two common materials used in H$\alpha$ narrow-band filters are SiO$_2$, which has $n=1.46$, and Ta$_2$O$_5$, which has $n\approx2$. Assuming a stack made of these materials,
we plot the theoretical shift in central wavelength of a 3nm bandpass narrow-band filter as a function of angle of incidence of incoming light (or tilt of the filter) in the left-hand panel of Figure~\ref{fig:tiltshift1}, for a filter with a central wavelength of 659.9 nm for light with 0$^{\circ}$ angle of incidence. Two tilts are marked with maroon circles corresponding to wavelength shifts that would bring the central wavelength to the rest frame wavelengths of [NII] and H$\alpha$. 
The left-hand panel of Figure~\ref{fig:tiltshift1} shows that over an angle of incidence of 20 degrees (which would be a modest angle of incidence range for a filter placed in the pupil of a reimaging camera on a large telescope) the central wavelength shifts by over 8 nm. Depending on the location of the filter, this effect is averaged over the spatial field of view, or over the pupil plane, or both. While the details depend on the specific filter design, the effect is generic, and it places severe limitations on the usefulness of very narrow bandpass filters on large telescopes.

\begin{figure} [ht]
   \begin{center}
   \begin{tabular}{c} 
     \includegraphics[height=10cm]{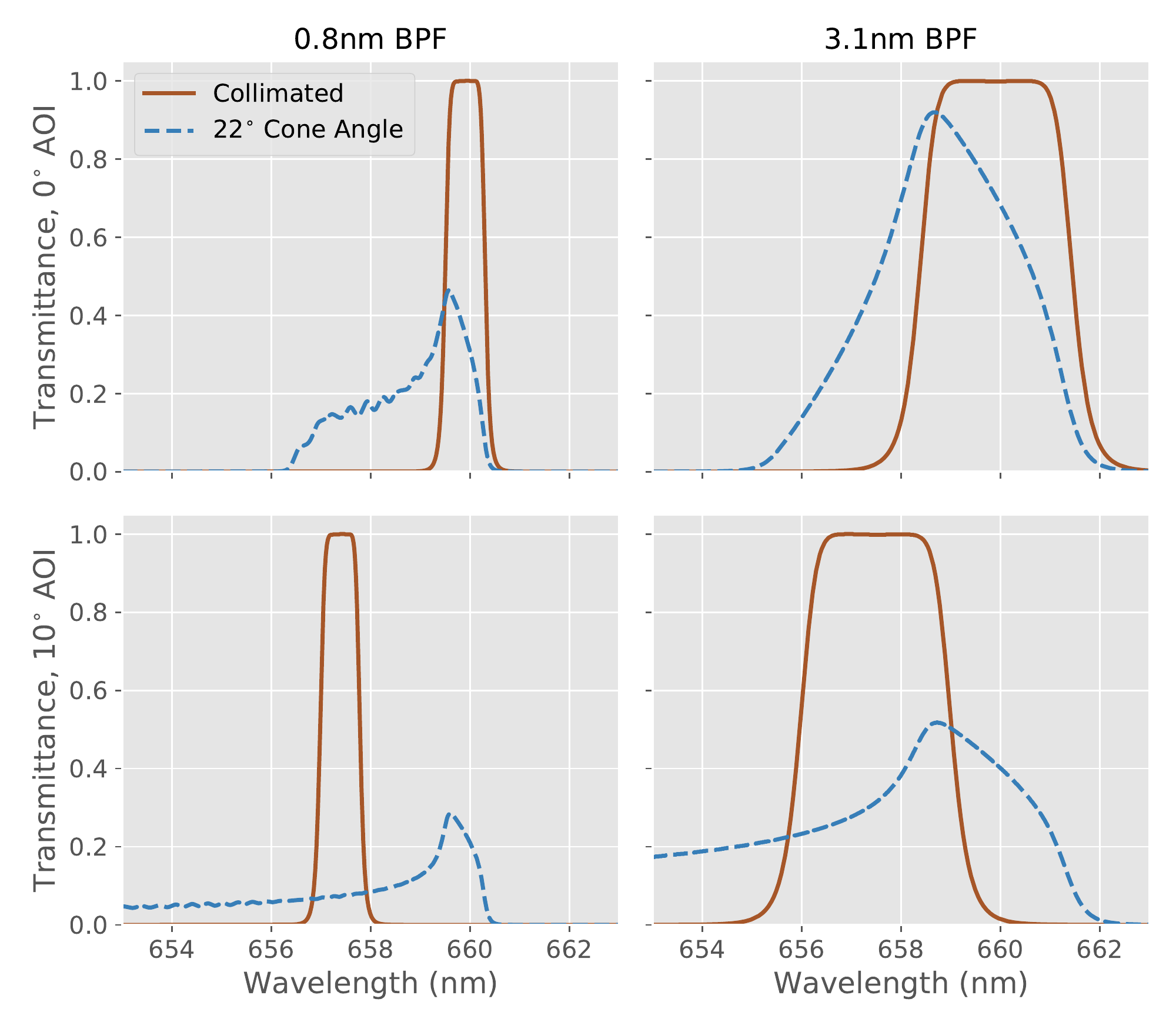}
   \end{tabular}
   \end{center}
   \caption[throughput] 
   { \label{fig:throughput} 
    The transmission through the narrow-band filters of collimated incident light (maroon solid curves) and where the incident light has a 22$^{\circ}$ cone angle (blue dashed lines). The top and bottoms panels show the transmittance at 0$^\circ$ and 10$^\circ$ angle of incidence, respectively. The left and right panels show the transmittance for 0.8 nm and 3.1 nm bandwidth filters, respectively. The throughput for incident light in a converging beam is degraded such that the effective bandwidth is much wider than the specifications, an effect that becomes more drastic as the bandwidth becomes narrower.
}
\end{figure} 
   
We believe that this `bug' limiting the performance of narrow band filters on large telescope can be turned into a `feature'. As the right-hand panel of Figure~\ref{fig:tiltshift1} shows, by placing the filter at the front of the system the angle of incidence variation across the pupil is always small, and by tilting a full-aperture filter by 7.5$^{\circ}$ one could (in this case) shift the central wavelength of the filter to target the wavelength of an [NII] emission line for an astronomical object, while a tilt of 12.5$^{\circ}$ would isolate H$\alpha$ emission line from that same source. In this example, a single narrow-band filter could be used to turn a Dragonfly-like system into a wide-field ultra-narrow band imager that isolates two particularly interesting emission lines.

How important is it that the tiltable filters be placed at the front of the optical system? It would be much simpler to tilt small filters near the back of the telescope than large filters at the front of the telescope. However, doing so strongly degrades the performance of the filters. At the front of the telescope the incoming wavefront is collimated and has a focal ratio of infinity. When placed near the sensor, the filters are inside a converging beam, with a cone angle of order tens of degrees for many practical setups.\footnote{In the specific case of the Dragonfly telescope, with individual cameras operating at $f/2.8$, the cone angle of light at the typical filter location (near the sensor) is around 22$^{\circ}$. In large reimaging cameras, the cone angles are generally steeper.}
In Figure~\ref{fig:throughput}
we compare the effective transmittance through a 3nm narrow-band filter stack where the incident light is collimated and where the incident light is arriving with a 22$^{\circ}$ cone angle, at two different tilts (0$^\circ$ and 10$^\circ$).\footnote{Data kindly provided by Iridian Spectral Technologies, based on their proprietary 3nm filter design. The general behaviour shown is generic for all filters of comparable bandwidth.} 
The effect of the wide cone angle is very large: the transmission function of the collimated light through the narrow-band filters at both tilts is essentially rectangular (consistent with the design), but the transmittance of the converging beam of light departs dramatically from the nominal transmission function, with peak transmission lower and and bandwith much wider. Even at zero tilt, the figure confirms the generally-held view that there is an effective lower limit on the full-width at half-maximum (FWHM) of a filter's transmission function in converging beams.  Extremely narrow bandpass interference filters ($<$10 nm FWHM) can be manufacured and used on large telescopes, but their transmittance is that of a filter much wider effective bandpass. Placing the narrow-band interference filters in a collimated beam, such as at the entrance pupil at the front of a telescope, mitigates this issue.
   
\section{The Dragonfly Filter-Tilter}

\subsection{Design}
\begin{figure} [ht]
   \begin{center}
   \begin{tabular}{c} 
   \includegraphics[height=4.5cm]{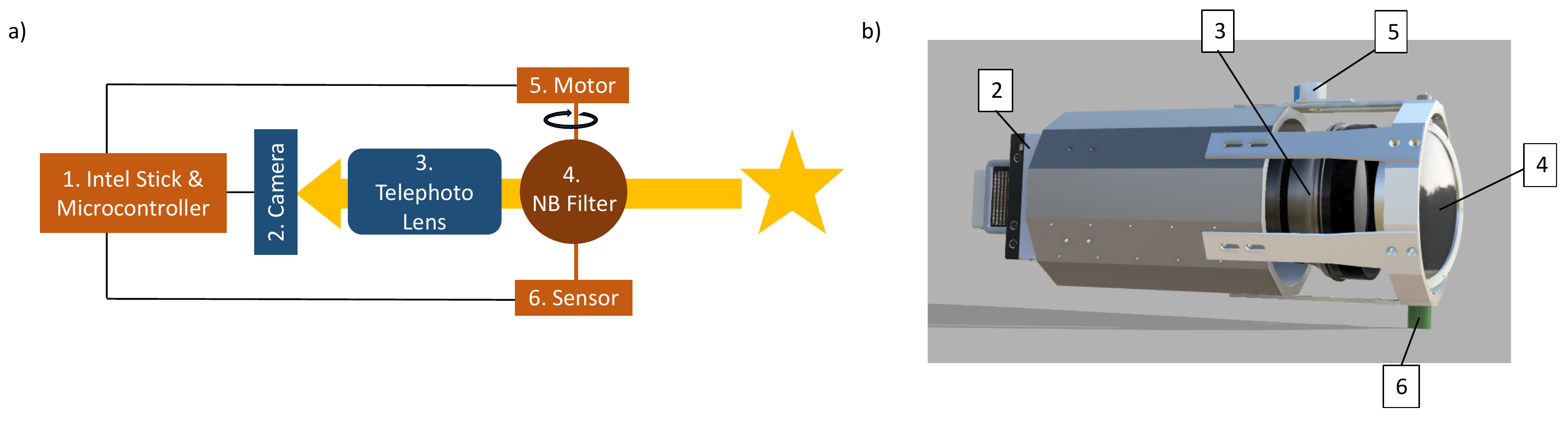}
   \end{tabular}
   \end{center}
   \caption[CAD] 
   { \label{fig:CAD} 
   Illustrations of a unit of the Dragonfly Telephoto Array with the Filter-Tilter instrumentation. 
   (a) A diagram of the optical path of Dragonfly, which is identical for each unit of the telescope. Incident light enters from the right of the diagram, passing through the narrow-band filter first (label: 4), then the lens (label: 3), and finally the camera (label: 2).  The Intel Stick (label: 1) is connected to the camera and receives the image.  The motor (label: 5) tilts the filter with respect to the optical axis; the tilt is read out by the angle sensor (label: 6). Both motor and sensor are controlled by a microcontroller (e.g. an Arduino; label: 1), which reports to the Intel Stick. A baffle to block out stray light is not shown. 
   (b) A CAD model of one unit of the Dragonfly Filter-Tilter instrumentation with the same parts as in (a) labelled for reference.  
}
\end{figure}

We have designed an instrument, the `Filter-Tilter', to test the ideas described in the previous section. Figure~\ref{fig:CAD} presents a block diagram (left) and a CAD model (right) of a Filter-Tilter affixed to a unit of Dragonfly telephoto array. Incident light enters from the right and passes through the narrow-band filter first, which is located at the entrance pupil of the optical system for each unit, then is received by the Dragonfly unit optical tube assembly (a Canon Telephoto lens and CCD camera, which is controlled by a microcomputer, currently an Intel Compute Stick).  The central axis of the filter is connected to a stepper motor, which rotates the filter with respect to the optical axis, and an angle sensor on the other, which reads the tilt of the filter. Both motor and sensor are connected to a microcontroller, which is in turn connected to the Intel Compute Stick. To tilt the filter, commands are sent from the microcomputer to the stepper motor,
and readings from the the angle sensor are used to close the feedback loop.

The filters in our prototype system have a bandwidth of 3 nm. They are manufactured by Iridian Spectral Technologies\footnote{https://www.iridian.ca/} and have excellent optical and spectral performance. The clear aperture of the filters is 150 mm, which allows the filters to be tilted to 20 degrees before vignetting the 142.5 mm clear aperture of the Canon EF 400 mm f/2.8 IS lenses in the Dragonfly array. Our optical setup feeds SBIG Aluma 694 CCD cameras manufactured by Diffraction Limited\footnote{https://diffractionlimited.com/}. These cameras use a Sony ICX-694 CCD sensor which was chosen on the basis of having both low dark current (0.025 electron s$^{-1}$ pixel$^{-1}$) and low read noise (4.5 electron RMS).

As shown in Figure~\ref{fig:CAD}, each filter is held in a circular cell (the `inner ring'), and this is encompassed by an `outer ring'. The inner ring tilts within the outer ring, riding on short pins which allow free movement. The outer ring is affixed by six legs to the supporting tube of each each Dragonfly element.  A stepper motor is mounted on one of the legs, and an angle sensor is mounted on the outer ring. All inner faces of the Filter-Tilter are anodized black to reduce reflections in the telescope. An outer baffle (not shown) is used to block stray light from entering the filter tilter. This baffle was 3D printed from ABS material with black optical flocking paper affixed to its interior to reduce stray light.

The stepper motor and angle sensor must each be less than 3~cm in height to avoid interference between lenses. For this reason, a small hobbyist stepper motor (28BJY-48) with 2048 steps per revolution (or 0.18 degrees per step) and dimensions of 28 mm x 19 mm was selected, primarily on the basis of its small size. For similar reasons, a hobbyist angle sensor (Gravity 360 Degree Hall Angle Sensor) was chosen. The sensor has a resolution of 0.088 degrees, which is sufficient for our prototype. Both the motor and the sensor interface to the microcomputer using a microcontroller (an Arduino Uno). The stock microcontroller's 10-bit analog-digital converter (ADC) cannot read the angle sensor with the needed precision, so an Arduino shield was built (using a custom printed circuit board) to provide a 12 bit ADC, which is used to read out the angle sensor. The microcomputer communicates with the microcontroller using serial commands.

\subsection{Tilting Precision \& Accuracy Requirements}
In order to accurately observe line emission at a specific wavelength, the tilting of the filter needs to be precise and accurate to some level of tolerance. We can find this level by first considering the filter response curve. With a 3 nm filter bandwidth, a tilting accuracy within $\pm1.5$ nm will achieve our desired goal (the line wavelength within the bandpass), assuming that the filter can be tilted with enough precision to reach the desired tilt. With the 12 bit ADC used to obtain the tilt from the angle sensor, the precision of the tilt is $\pm0.088^{\circ}$. The accuracy of the angle sensor is 0.3\% of its full scale, which yields $\pm1.08^{\circ}$.  The total uncertainty in the central wavelength resulting from the angle sensor increases with the tilt of the filter so taking the maximum tilt of 20$^{\circ}$, the uncertainty from the angle sensor has a maximum of $\pm0.9$ nm, which is within the stated tolerance of $\pm$1.5 nm for the 3 nm bandwidth. On the other hand, with a filter bandwidth is 0.8 nm, the tilting tolerance is $\pm0.4$, which means that the current angle sensors are not accurate enough for the narrower bandpass\footnote{Alternative methods for measuring the filter tilt, such as using incremental encoders, are being investigated to increase the tilting accuracy for future design iterations.}.

\begin{figure} [ht]
   \begin{center}
   \begin{tabular}{c} 
   \includegraphics[height=6.5cm]{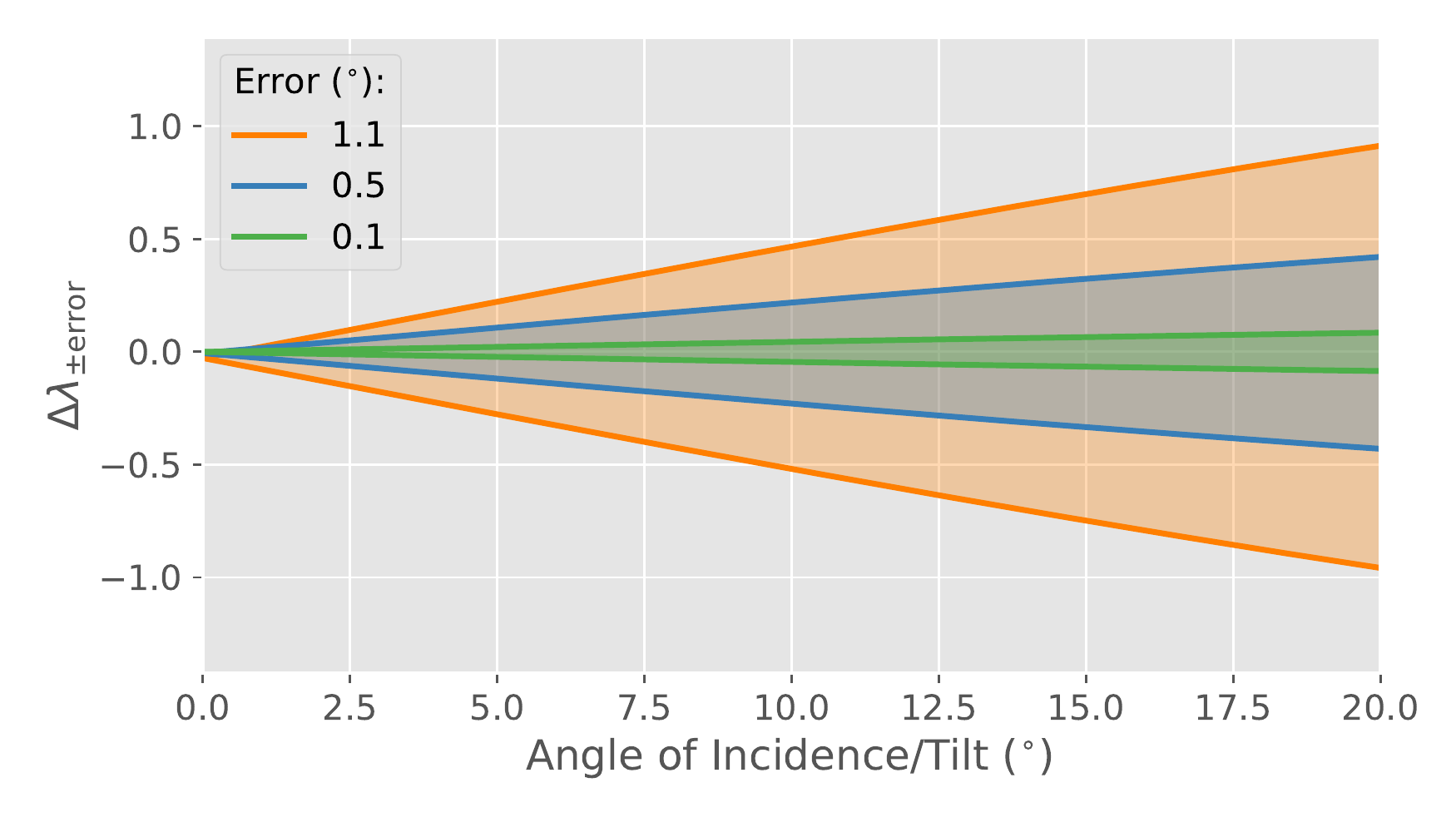}
   \end{tabular}
   \end{center}
   \caption[precision] 
   { \label{fig:precision} 
   The uncertainty of the filter transmission wavelength due to the uncertainty of the tilting, shown for specific uncertainties of 0.1, 0.5, and 1.1 $^{\circ}$.  The uncertainty in the filter transmission wavelength increases as the tilt of the filter increases, due to the response of the CWL with filter tilt.
}
\end{figure} 

One additional consideration is the presence of sky lines near the emission line of interest, which would place tighter requirements on the tilting accuracy in order to avoid including the sky line within the bandpass during observations. From current limitations on low surface brightness measurements, this may prove necessary to reach the surface brightness limits required for observing the emission from gas in the circumgalactic medium\cite{fuma17}. In Figure~\ref{fig:precision}, we explicitly demonstrate the resulting uncertainty in central wavelength as a function of tilt. For example, with a tilting uncertainty of $1.1^{\circ}$, the uncertainty of the filter central wavelength ranges from $\pm 0.25$ nm to $\pm 0.9$ nm  between tilts of 5$^{\circ}$ to 20$^{\circ}$ (the orange shaded area).
One can think of the boundaries of the shaded areas in Figure~\ref{fig:precision} as the minimum allowable separation between the line of interest and a sky line for each stated uncertainty in tilting of the filter. The current angle sensor, with its uncertainty of $1.08^{\circ}$, corresponds to the orange area in Figure~\ref{fig:precision}. Reducing the uncertainty in tilting (e.g. to 0.5$^{\circ}$ or 0.1$^{\circ}$) will allow us to image emission lines that fall nearer to sky lines.


\section{Laboratory Testing}
\label{sec:Tests}

   \begin{figure} [ht]
   \begin{center}
   \begin{tabular}{c} 
   \includegraphics[height=8.5cm]{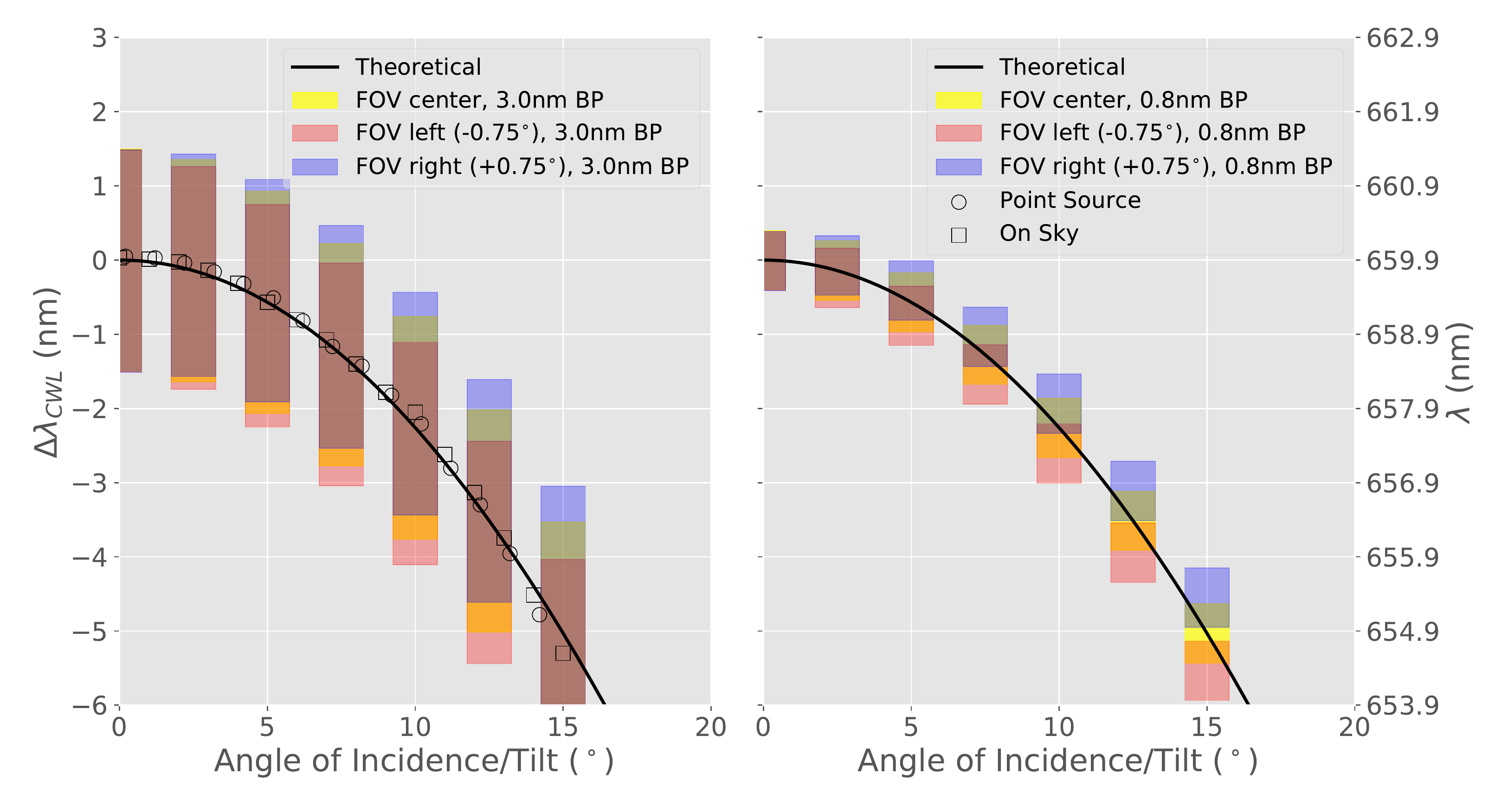}
   \end{tabular}
   \end{center}
   \caption[tiltshift] { \label{fig:tiltshift}  The theoretical shift of the central wavelength of the filter as a function of angle of incidence (or, equivalently, the tilt of the filter with respect to normal incidence) is shown as the black line on the left and right panels.  The filter bandpass at a series of tilts and as seen by light incident from the left (center; right) of the field-of-view is represented by the pink (yellow; blue) rectangular tiles: each tile has a width corresponding to the field-of-view (in degrees) and a height corresponding to the filter bandpass (in nm).  In the left (right) panel, the bandwidth of the filter is 3.0 (0.8) nm.  In both panels, the field-of-view of the filter is taken as 1.5$^{\circ}$ to represent the current CCD cameras (this is the smaller axis of the field-of-view).  As the filter tilts, the effective bandpass for light from either side of the FOV separates further.}
   \end{figure}

The performance of a prototype Filter-Tilter was evaluated in our laboratory
at the University of Toronto. We measured the throughput of the filters as a function of angle of incidence (or the tilt of the filters, equivalently).  
The test set-up consisted of a low resolution spectrometer (an Alpy 600 manufactured by Shelyak LLC\footnote{Shelyak Instruments} with a spectral resolving power of R~$\sim$~600) attached to a Canon EF 400 mm f/2.8 IS II USM lens placed inside a Dragonfly aluminum tube with a Filter-Tilter attached. The spectrometer's slit was
located at the infinity focus position of the lens. Spectra were taken at various tilts of the Filter-Tilter using both a white light lamp (with a collimating lens to mimic a source at infinity) and the Sun as sources.\footnote{The white light source was aligned with the optical axis by imaging with a CCD camera prior to attaching the spectrometer. A tilt of zero degrees can then be established easily by autocollimation. At zero degrees tilt, the reflected spot from the filter falls on the light source.} 

The left-hand panel of Figure~\ref{fig:tiltshift} shows our laboratory-based measurements of the central wavelength of the filter transmission as a function of filter tilt. Circles and squares correspond to measurements made using the lamp and the Sun, respectively. The solid line in this panel shows the predicted performance of the filter, based on Equation (1) with an effective index of refraction computed by fitting to the data. Our conclusion from these measurements is that the assumptions underlying Equation (1) provide a good description of the physical situation.

\section{Additional Considerations}

Optimal application of the Filter-Tilter requires that some special attention be given to flat fielding, because the tilted element in the Filter-Tilter breaks the azimuthal symmetry of the optical system, resulting in a slight gradient in the central wavelength across the field of view. This can be calibrated out using a tilt-dependent illumination correction. Another practical consideration springs from the fact that integrations made using narrow bandwidth filters may not be sky noise limited, unless detectors with extraordinarily low read noise are used, or the data are obtained using unusually long integration times. In this section we will sketch these considerations, deferring detailed discussion to a future paper (Lokhorst et al. 2021, in preparation) which will present on-sky data obtained with the prototype array.

\subsection{Tilt-dependent bandpass across the field of view}\label{sec:debeffect}

\begin{figure} [ht]
   \begin{center}
   \begin{tabular}{c} 
   \includegraphics[height=10cm]{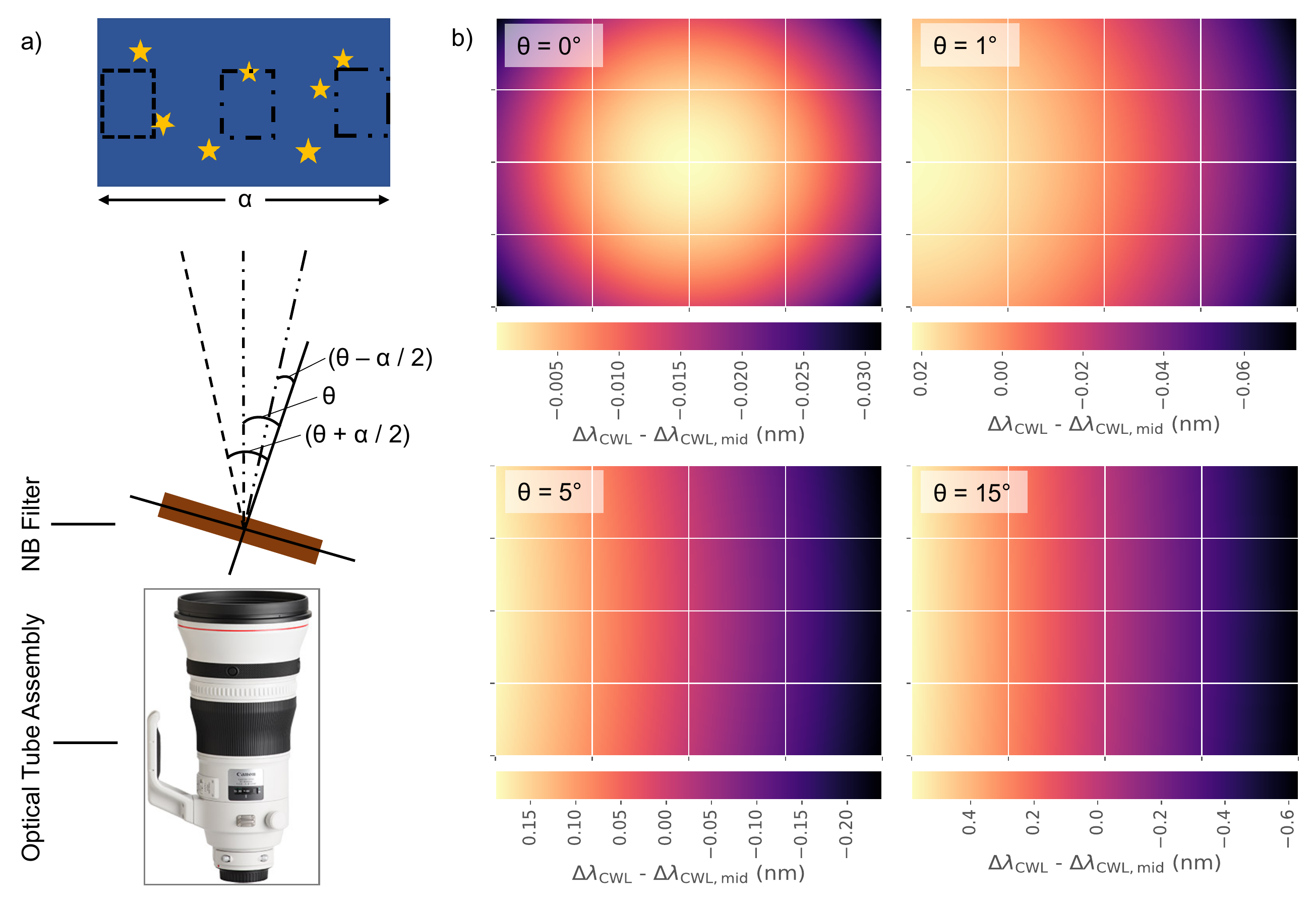}
   \end{tabular}
   \end{center}
   \caption[debeffect] 
   { \label{fig:debeffect} 
   (a) An illustration of the angle of incidence for sources of light at each side and the center of the field-of-view.  At each side of the field-of-view, the light arrives on the filter at an angle equal to the angular distance from the light source to the center of the field-of-view.
   (b) The central wavelength of the narrowband filter across the field-of-view of the CCD camera (1.5$^{\circ}\times1.9^{\circ}$). The effective ``tilt" seen by the incident light is the combination of the angle-of-incidence of light arriving from across the field-of-view and the tilt of the filter.  The color map represents the difference between the actual filter central wavelength ($\Delta\lambda_{\mathrm{CWL}}$) and the expected central wavelength of the filter (which occurs at the middle of the field-of-view; $\Delta\lambda_{\mathrm{CWL, mid}}$) across the field-of-view for the current cameras at four different filter tilts.  The leftmost panel shows the filter central wavelengths with the filters at zero degree tilt: the response is symmetric across the field-of-view, as expected, since the filter central wavelength is a symmetric function about zero degrees of tilt. As the tilt is increased (to 1$^{\circ}$, 5$^{\circ}$, and 15$^{\circ}$ for the mid-left, mid-right, and rightmost panels, respectively), the the symmetry drops off as the response leaves the peak of the shift-tilt relation around zero degree tilt.
}
   \end{figure}

An important point that we have yet to consider is the extent to which a filter that is rotated to be normal to the optical axis ({\em i.e.} at a zero degree mechanical tilt) is still seen as `tilted' by wavefronts coming in at varying field angles. The `effective' tilt experienced by incident light is the combination of the angle of incidence of light across the field-of-view with the mechanical tilt of the filter (this is illustrated in the left-hand panel of Figure~\ref{fig:debeffect}). Equation (1) can be used to understand the size of this effect, which manifests itself as a shift in the central wavelength of the filters as a function of position across the field of view, and which is quantified in Figure~\ref{fig:tiltshift} by colored tiles. These tiles correspond to the FWHM of the filters at various tilts and field angles. The two panels in the Figure correspond to filters with bandpasses of 3.0 degrees (left-hand panel) and 0.8 degrees (right-hand panel). With a 3 nm bandpass filter over the field of view of the Dragonfly Filter-Tilter experiment (1.5$^{\circ}\times1.9^{\circ}$), the entire field of view is essentially monochromatic for all tilts considered (up to about 20$^{\circ}$). However with a 0.8 nm bandpass filter, the yellow and blue tiles (representing the bandpass of the filter at either side of the field-of-view, respectively) have no overlap when the tilt is greater than $\approx12.5^{\circ}$. For the 0.8 nm bandwidth filter, the monochromatic field-of-view at a tilt of 20$^{\circ}$ is $\lesssim$1$^{\circ}$. 
We explicitly model this effect across the field-of-view for four different mechanical filter tilts.  Maps of the difference between the effective filter central wavelength ($\Delta\lambda_{\mathrm{CWL}}$) and the expected central wavelength of the filter (which occurs at the middle of the field-of-view; $\Delta\lambda_{\mathrm{CWL, mid}}$) are shown in the right panel of Figure~\ref{fig:debeffect}.  At zero degree tilt, the response is symmetric across the field-of-view since the filter central wavelength is a symmetric function about zero degrees. As the tilt is increased to 1$^{\circ}$, 5$^{\circ}$, and 15$^{\circ}$, the response is no longer symmetric, as the response of the filter with angle of incidence leaves the peak of the shift-tilt relation. As the tilt increases, the variation in central wavelength becomes larger: at 15$^{\circ}$ tilt, the variation across the field-of-view is $\sim$1.1 nm, which is 36\% the size of the 3nm bandwidth and 140\% the size of the 0.8 nm bandwidth. When designing a survey, this effect must be taken into account as it can limit the usable field-of-view or require additional observing times at subsequent tilts. Conversely, this effect could be possibly be useful for specific targets where emission is Doppler shifted across the field-of-view, such as large scale galactic outflows or inflows. 


One final note is that these models demonstrate that there will be a difference in the response across the field-of-view depending on the orientation of the filter with respect to the field-of-view.  For example, if the tilting axis of the filter is flipped from EW to WE (consider flipping the panels in Figure~\ref{fig:debeffect} about a vertical axis), the effective central wavelength at one side of the field-of-view could be changed by up to $\approx$1 nm. 
The change of incident angle is straightforward to model as shown, thus the response of the filter can be modelled across the filter allowing observations to be corrected, but requires additional considerations when planning and carrying out observations. Furthermore, if the sensitivity of the pixels in the sensor has a significant chromatic dependence over very short wavelength baselines, the effect may result in flat field non-uniformity varying as a function of tilt. To ameliorate this, the prototype array employs deployable luminescent flat field panels on individual lenses, which allow flats to be obtained before and after science data is taken.

\subsection{Special challenges for low surface brightness imaging}\label{sec:noise}

\begin{table*}[ht]
\caption{Noise sources and detection limits for the Dragonfly system. 
The noise components were calculated assuming an exposure time of 30 minutes and a system efficiency of $\epsilon$=0.64 (assuming throughput of filters is 100\% for now, and measured camera quantum efficiency of 0.75 and lens throughput of 0.85).  One column shows the cumulative noise within one pixel (2.45'') and the other shows noise within a binned 4x4 pixel area, which corresponds to a 10 arcsec scale.} 
\label{tab:NoiseSources}
\begin{center}       
\begin{threeparttable}
\begin{tabular}{ *4l }    
\toprule
Contaminant & Per Pixel, Per Unit  & 4x4 Pixels, 100 Units &   \\\midrule
Sky signal\tnote{a} &  1.5 (0.4) ph s$^{-1}$ arcsec$^{-2}$ m$^{-2}$  & - & \\
Dark current & 0.025 electron s$^{-1}$   & - & \\
Read out noise  & 4.5 electron & -  &   \\
\\
Contaminant \emph{per exposure} &   \\\midrule
Sky signal\tnote{a}  &  170 (42) e- & 1.7$\times10^{5}$ (4.2$\times10^{4}$) e- &   \\
Dark current  & 45.0 e-  & 4.5$\times10^{4}$ e- &  \\
Sky shot noise\tnote{a} & 13.0 (6.5) e- & 4.1$\times10^{2}$ (2.1$\times10^{2}$) e- &    \\
Dark current noise & 6.7 e-  & 2.1$\times10^{2}$ e- &  \\\\
\toprule

1$\sigma$ Detection Limit (30 min)\tnote{a} 
 & 0.14 (0.092) & 3.3$\times10^{-3}$ (2.2$\times10^{-3}$) \\
 & ph s$^{-1}$ arcsec$^{-2}$ m$^{-2}$ & ph s$^{-1}$ arcsec$^{-2}$ m$^{-2}$ \\
 & \emph{-or-} & \emph{-or-} \\
 & 600000 (390000) ph s$^{-1}$ sr$^{-1}$ cm$^{-2}$ & 14000 (9200) ph s$^{-1}$ sr$^{-1}$ cm$^{-2}$ \\
\\
1$\sigma$ Detection Limit (100 hr)\tnote{a,b}
 & 0.0095 (0.0062) & 2.4$\times10^{-4}$ (1.6$\times10^{-4}$) \\
 & ph s$^{-1}$ arcsec$^{-2}$ m$^{-2}$ & ph s$^{-1}$ arcsec$^{-2}$ m$^{-2}$ \\
 & \emph{-or-} & \emph{-or-} \\
 & 40300 (26200) ph s$^{-1}$ sr$^{-1}$ cm$^{-2}$ & 1000 (660) ph s$^{-1}$ sr$^{-1}$ cm$^{-2}$ \\

\bottomrule
\end{tabular}
  \begin{tablenotes}
  \item[a] Calculated respectively for the 3 (0.8) nm bandwidth filter.
  \item[b] Total of 200 30-minute exposures.
  \end{tablenotes}
\end{threeparttable}
\end{center}
\end{table*}

The Dragonfly Telephoto Array's fundamental goal is to probe down to extremely low surface brightness levels. Because of the narrowness of the spectral bandpass, it is difficult to be sky noise limited when imaging through filters with $\sim 1$~nm bandpass,
at least using the `prosumer' CCD cameras presently used on the array. The main sources of noise in our data are summarized in Table~\ref{tab:NoiseSources}. 
Approximate values for the sky signal observed within our instrument are calculated by integrating the 3 nm and 0.8 nm narrowband filter transmission curves over the sky background spectrum\footnote{https://www.eso.org/observing/dfo/quality/UVES/pipeline/sky\_spectrum.html} taken by the Ultraviolet and Visual Echelle Spectrograph on VLT. The other major sources of noise are from the CCD detector: the read-out noise (4.5 electron RMS) and dark current (0.025 e- s$^{-1}$ pixel$^{-1}$). 
Long exposure times ($\sim$30 minutes) are needed for sky noise to be the main source of random error.
With such long exposure times, dark current becomes comparable to the read-out noise and eventually dark current could become limiting sources of noise in our data.
The important lesson here is that extreme narrow band imaging places quite stringent requirements on the properties of the detectors. For these reasons we are
investigating new camera designs which lower the operating temperature of the 
detectors, and new sensor technologies (such as CMOS) which would, in principle,
lower the read noise. In any case, for the Dragonfly mosaic design where data is stacked from each lens to form a final deep image, each frame needs to be shot-noise limited to preserve the signal-to-noise gain when stacking the images. 


A great benefit of mosaic telescope arrays is that their effective aperture can be scaled up by adding more telescopes to the system. In Table~\ref{tab:NoiseSources} we give the 1$\sigma$ detection limits for a single unit and for a 100-unit array. To bracket the capabilities of these systems, numbers are shown assuming a single 30-minute integration with the final data unbinned, and for a 100 hour total exposure time binned to 10 arcsec scales. The sensitivity has a straightforward scaling with the number of units and total exposure time, and a limiting surface brightness of 1000 ph s$^{-1}$ sr$^{-1}$ cm$^{-2}$ is reached with a 100-unit system in 100 hours of integration, assuming the data is binned to 10 arcsec scales. These detection limits are an order of magnitude fainter than those reached by typical current wide-field narrowband imaging (e.g. Refs.~\citenum{dona95,hlav11,gava17}).


\section{Summary}

We present a novel technique for wide-field ultra-narrow bandpass imaging that holds some promise for allowing line emission from astrophysical phenomena to be probed down to very low surface brightness levels. 
Our approach centers upon the idea of locating tiltable $\sim 1$~nm bandpass narrow band filters at the entrance pupil of a telescope.  In this paper we describe a specific implementation of this idea for the Dragonfly Telephoto Array: the `Filter-Tilter'. 
The Filter-Tilter has been tested in our optical laboratory at the University of Toronto, and in this paper we present measurements of the spectral throughput of the instrument as a function of the tilt of the filter. These measurements confirm the theoretical relationship between the angle-of-incidence of light on the filter and the resulting shift in the central wavelength of the filter. 

In addition to describing the Filter-Tilter, we highlight some considerations that are relevant for planning and carrying out ultra-narrow band observations. These include:

\begin{enumerate}
    \item The effect of the range of incident angles from the field-of-view, which will result in a relative shift of the central wavelength across the field-of-view. The effective filter central wavelength across the field-of-view is easily modeled, and in cases where this shift in central wavelength matters, it can be corrected for
    fairly easily. 
    
    \item The sensitivity and detection limits for the Dragonfly Filter-Tilter, and its dependence on the number of lenses in the array, binning factor, and exposure time. We show that reaching a surface brightness limit of below $\sim$1000 ph s$^{-1}$ cm$^{-2}$ sr$^{-1}$ is achievable. This is an order of magnitude deeper than the present limit with conventional wide-field survey telescopes.

    \item As filter bandpasses are narrowed, camera read noise and dark current begin to become important factors in the error budget. The current cameras on the Dragonfly Telephoto Array meet our requirements with 3~nm bandpass filters, but future surveys to be undertaken using sub-nm bandpass filters would benefit from being done with cameras having lower read noise. Suitable cameras exist, and we have begun testing them in our lab.
\end{enumerate}

\noindent A prototype three-element Dragonfly Filter-Tilter array is currently on-sky and taking data. First results will be presented in a companion paper (Lokhorst et al. 2021, in preparation).




\acknowledgments
 

We are thankful for contributions from the Dunlap Institute (funded through an
endowment established by the David Dunlap family and the University of Toronto) which made this research possible.
The SBIG Aluma 694 cameras on the prototype Dragonfly array are manufactured by Diffraction Limited in Ottawa, Ontario. We thank Doug George and Colin Haig at Diffraction Limited for helping us take advantage of their excellent hardware. The narrow-band filters in our prototype array are manufactured by Iridian Spectral Technologies in Ottawa, Ontario.
Thanks are due to Jason Padiwar and his colleagues at Iridian Spectral Technologies who worked hard to design, manufacture and characterize our ultra-narrow bandpass filters.

This research made use of Astropy,\footnote{http://www.astropy.org} a community-developed core Python package for Astronomy \cite{astropy:2013, astropy:2018}.

We acknowledge the support of the Natural Sciences and Engineering Research Council of Canada (NSERC).

Nous remercions le Conseil de recherches en sciences naturelles et en g\'enie du Canada (CRSNG) de son soutien.


\bibliographystyle{spiebib} 

\end{document}